\documentclass[conference]{IEEEtran}
\IEEEoverridecommandlockouts
\usepackage{cite}
\usepackage{graphicx}
\usepackage{amsmath}
\usepackage{subfigure}
\usepackage{gensymb}
\usepackage{xcolor}
\usepackage{lipsum}
\usepackage{multirow}

\begin{document}
\title{Reflection Channel Model for Terahertz Communications\\
\thanks{This work was funded by the NSF under grant CNS-1910655.}
\thanks{All our data and code is available at: http://web.cecs.pdx.edu/~singh/projects/thzmimo/Home.html}
}

\author{\IEEEauthorblockN{Thanh Le}
\IEEEauthorblockA{Department of ECE\\
Portland State University\\
Portland, OR 97207\\
Email: thanh4@pdx.edu}
\and

\IEEEauthorblockN{Ha Tran}
\IEEEauthorblockA{Department of ECE\\
Portland State University\\
Portland, OR 97207\\
Email: tranha@pdx.edu}
\and

\IEEEauthorblockN{Suresh Singh}
\IEEEauthorblockA{Department of Computer Science\\
Portland State University\\
Portland, OR 97207\\
Email: singh@cs.pdx.edu}}
%
%
%


{}



\maketitle

\begin{abstract}
 Terahertz frequencies are an untapped resource for providing high-speed short-range communications. As a result, it is of interest to study the propagation characteristics of terahertz waves and to develop channel models. In previous work we used a measurement-based approach to develop an accurate channel model for line of sight (LoS) links. In this paper we extend that work by developing channel models for non-line of sight (NLoS) links where the signal suffers one reflection. We study reflections that occur off a metal plate as well as a piece of wood.
 
 Our model for received magnitude includes the effects of standing waves that develop between the transmitter and receiver. Measurements show an excellent agreement between empirical data and the model. In addition, we have analyzed the received phase of the reflected signal at frequencies in the range 320-480 GHz. We observed a linear error between the predicted and actual phase and developed a model to accommodate that discrepancy. The final model we have developed for predicting received phase is very accurate for the entire range 320 - 480 GHz and for both materials.
\end{abstract}


\section{Introduction}

The terahertz frequency band, which extends from 100 GHz to 3 THz is being studied for the next generation of wireless communication systems \cite{hemadeh18cst,xing19globecom}. Expected applications include ultra high data rate short-range communication (e.g., device to device or chip-to-chip), as an alternative to free-space optics, and as a wifi-like system for small offices etc. There are many challenges in deploying terahertz communication systems including its high attenuation with distance as well as with reflections and frequency-specific molecular absorption. Among the frequencies available, channels at 140, 220, 340, 410, and 460 GHz have been identified as viable communication channels as they suffer the least (though still considerable) environmental attenuation \cite{john10jap}. Thus, several different research efforts have been directed at developing communication systems at these bands. Examples of work most related to ours include prototype communication systems as proof of concept \cite{kallfass2011, Hirata2010, Koenig2013, Song2012}, and channel models for line of sight (LoS) communication \cite{kim16twc,han15twc,saleh87jsac,hemadeh18cst,tranEffectStandingWave2021}. Our work has been on using measurements to develop channel models for LoS channels and MIMO LoS channels \cite{tranEffectStandingWave2021,moshirModulationRateAdaptation2016,singhChallengesTerahertzMIMO2019,singhMeasurement2x2Terahertz2020}. In this paper we extend the channel model for the case of terahertz reflections off {\em metallic} and {\em wooden} surfaces. The channel model shows good agreement with measurements for both cases.

The usage scenario for terahertz reflections is best explained by considering an office with a ceiling mounted THz access point.
Because terahertz channels are primarily LoS, carefully planned reflections are needed in order to provide full coverage in these scenarios. For instance, due to environmental clutter such as furniture, direct LoS paths between the access point and a user may be blocked. Thus, to reach the user, a reflection off a ceiling mounted metal plate could be used. These types of surfaces are called passive reflective surfaces (as opposed to active reflective surfaces) and are inexpensive to use. Similarly, if there is a sufficient density of such reflective surfaces, we can envision multiple reflected paths existing between the AP and user which would allow us to provide a high data rate channel. Our measurements show that reflections off metal and wood provide a viable alternative signal path. Of course, as expected, reflections off wood are suffer much greater attenuation (approximately 10dB greater path loss as compared to metal).

The remainder of the paper is organized as follows:
\begin{itemize}
    \item In the next section we provide an overview of related work on channel modeling for terahertz,
    \item Subsequently in section \ref{testbed} we describe our experimental setup used for measuring reflections,
    \item Section \ref{model} presents our channel model and algorithm for fitting parameters,
    \item The model is validated against the measured data in section \ref{validate} and we conclude in section \ref{conclude}.
\end{itemize}

\section{Related Work} \label{related}

There has been considerable work on developing channel models for terahertz frequencies over the past few years such as ray tracing \cite{kurnerMeasurementsModelingBasic2015, hanMultiRayChannelModeling2015, gougeonRaybasedDeterministicChannel2019}, empirical fitting using free-space models \cite{sunPropagationModelsPerformance2018, sunInvestigationPredictionAccuracy2016, razavianTerahertzChannelCharacterization2021} and others \cite{juMillimeterWaveSubTerahertz2021,  yamamotoPathLossPredictionModels2008, guanMeasurementSimulationCharacterization2019, eckhardtChannelMeasurementsModeling2021, jornetChannelModelingCapacity2011, chengCharacterizationPropagationPhenomena2020}. These models typically have to account for the specific impairments suffered by terahertz radiation such as high path loss \cite{xingPropagationMeasurementSystem2018}, atmospheric absorption \cite{razavianTerahertzChannelCharacterization2021}, reflection from indoor building material \cite{jansenImpactReflectionsStratified2008, xingIndoorWirelessChannel2019}, time dispersion \cite{deng28GHz732015}, and RF impairments \cite{bouhlelSubTerahertzWirelessSystem2021}. Other papers have reported on measurements and modeling for specific use scenarios. For instance, \cite{razavianTerahertzChannelCharacterization2021} reports on long range (110m) terahertz links upto 1 THz, \cite{yamamotoPathLossPredictionModels2008,eckhardtChannelMeasurementsModeling2021} studies LoS and NLoS channels around vehicles, or train-to-infrastructure \cite{guanMeasurementSimulationCharacterization2019}, rack-to-rack and blade-to-blade communication in data centers \cite{chengCharacterizationPropagationPhenomena2020}, and device-to-device communication \cite{kurnerMeasurementsModelingBasic2015}.

In our previous work, we have used both a time-domain system (with HDPE lenses) as well as a VNA system to measure and subsequently characterize the channel \cite{singhChallengesTerahertzMIMO2019, singhMeasurement2x2Terahertz2020,moshirModulationRateAdaptation2016}. More recently, we observed that at specific distances the received power increases as compared with somewhat shorter distances. This observation led us to the discovery that {\em standing waves} produced between the transmit and receive antenna are responsible for the behavior \cite{tranEffectStandingWave2021}.

In this paper we extend our prior work to consider single reflections. There has been work done on terahertz reflections such as vehicle reflections \cite{eckhardtChannelMeasurementsModeling2021}, indoor materials including drywall, clear glass as well as stratified building materials \cite{xingIndoorWirelessChannel2019, maInvitedArticleChannel2018,jansenImpactReflectionsStratified2008}, and different metal materials \cite{maInvitedArticleChannel2018}. However, most of the work doesn't consider the effect of standing waves which is considerable in short-range communication. Our work in this paper will provide a more complete model that incorporates the effects of the standing wave and that predicts phase accurately.

\section{Measurement Setup and Experimental Design} \label{testbed}

A typical indoor signal path for sub-mmwave frequencies between transmitter and receiver is characterized by a line of sight path as well as numerous reflected paths which enables good coverage. At terahertz frequencies, on the other hand, most non-line of sight paths are absent due to air and material attenuation or direct blockage by obstacles. Thus, our goal is to determine if reflected paths can be created at these frequencies to enhance indoor coverage. Metals are ideally suited to this task because reflections are barely attenuated and indeed they act as mirrors at terahertz. We envision having metal sheets embedded in the ceiling of rooms to create a path between terahertz access points and devices that are not in the line of sight of the AP. {\em Our goal in this paper is to precisely characterize such a reflected channel}.

\begin{figure}[th]
  \centerline
      {\includegraphics[width=2.5in]{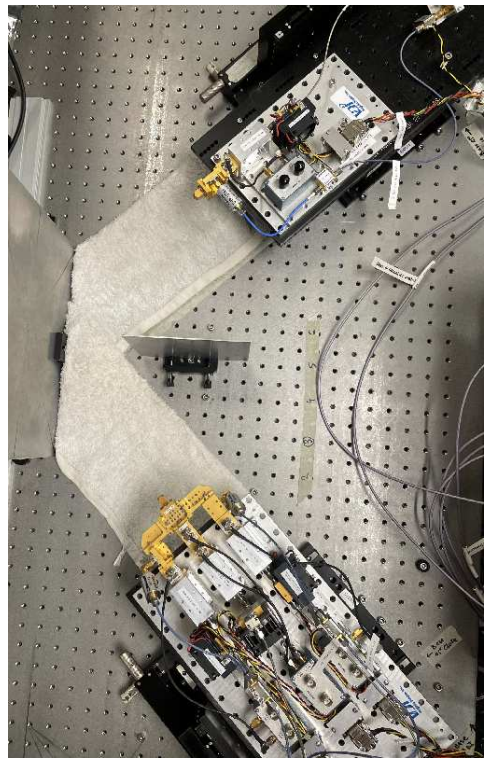}}
      \caption{Setup for reflection measurement at $45^o$ angle of incident}
      \label{setup}
\end{figure}

The measurement setup is illustrated in Figure \ref{setup} in which an aluminium plate is used to provide a NLoS signal path.  To ensure that there is no additional reflection path other than from the aluminium plate, wet towels are placed under the propagation path. We have previously shown that the inclusion of this wet towel eliminates the reflection from the optical table \cite{tranEffectStandingWave2021}. Another small aluminum piece is put in between the Tx and Rx to block any direct signals. An identical setup is used with wooden boards to characterize reflections off wood.
 
 To conduct the experiments, we use a Rohde \& Schwartz Vector Network Analyzer (VNA) and Virginia Diodes, Inc. (VDI) frequency extender modules. The WR2.2 band extender module is capable of generating signals from 325 GHz to 500 GHz. Before measurement, the system is calibrated using Thru-Reflect-Line (TRL) standards. This calibration moves the measurement reference plane to the waveguide opening. Other VNA settings are listed in Table \ref{table1}. For this band, horn antennas with 25dB gain are used to transmit and receive signals. The length of these antennas is 1.68 cm.
 
\begin{figure}[th]
  \centerline
      {\includegraphics[width=2.25in]{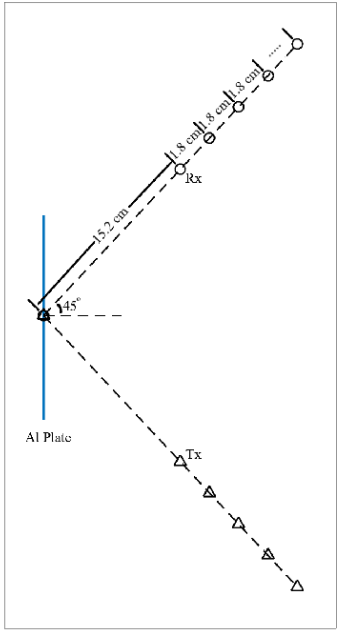}}
      \caption{Setup diagram with minimum distance and steps}
      \label{setupDiagram}
\end{figure}

 In this paper, we explore the case of $45^0$ incident angle. The minimum distance between the Tx/Rx to the Aluminium plate is 15.2 cm. For each measurement, both Tx and Rx are moved 1.8 cm further away from the Aluminium plate. Figure \ref{setupDiagram} shows the detailed diagram of the measurement setup with a total of 12 distances. Note that all distances are measured from one antenna's aperture to the other's.

\medskip
\begin{table}
\centering
\caption{VNA setting for WR2.2 band}
\begin{tabular}{|l|l|} \hline
Output power & 5 dBm\\
Frequency range & 325 - 500 GHz\\
Frequency step & 0.1 GHz\\
IF Bandwidth & 1 kHz\\
Averaging & 10\\ 
Calibration & Thru-Reflect-Line \\ \hline
\end{tabular}
\label{table1}
\end{table}
\medskip
 
\section{Channel Model} \label{model}

 For magnitude path loss fitting, we use the channel model from the previous work \cite{tranEffectStandingWave2021}. In this model, the standing wave effect between the transmit and receive horn antennas is taken into account to modify the {\em single-frequency floating intercept model} which has the path loss (PL) written as equation \eqref{sffl}.
 
\begin{equation}
\mbox{PL}(d) = \alpha + 10\beta\log_{10}\left(\frac{d}{d_0}\right) + X_\sigma, \mbox{    } d \geq d_0
\label{sffl}
\end{equation}
where $d$ is the distance, $d_0 = 15.2*2 = 30.4$ cm is the minimum distance, $\alpha$ is a floating intercept in dB that denotes the free-space path loss at $d_0$, $\beta$ is the distance-based path loss exponent, and $X_\sigma$ is the large-scale shadow fading that is modeled as a zero-mean Gaussian with standard deviation $\sigma$. 
 
\begin{figure}[th]
\centerline
{\includegraphics[width=1.75in]{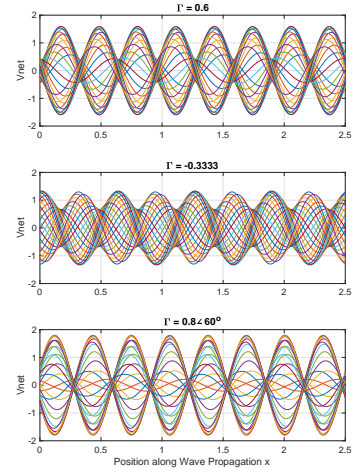}}
\caption{Example of standing waves.}
\label{standingwave}
\end{figure}

Equation \eqref{vnet-amplitude} shows the magnitude of the standing wave with $k$ as the wavenumber, and $\Gamma$ as the reflection coefficient. $A$ is complex amplitude corresponding to the forward wave at $d_0$.
\begin{equation}
|V_{net}(d)|^2 = \left[1 + |\Gamma|^2 + 2\Re(\Gamma e^{i2k(d-d_0)})\right] |A|^2\\
\label{vnet-amplitude}
\end{equation}
Figure \ref{standingwave} illustrates the effect of reflection coefficient $\Gamma$, wavelength $\lambda$ on the standing wave.

\subsection{Parameter Estimation for Magnitude} \label{magnitude}

We first fit the measured data into equation \eqref{sffl} to estimate the parameters $\alpha$ and $\beta$ using the least squares linear fit such that the root mean square deviation from the mean path loss is minimized. The difference between the fit and measured data is used to obtain the initial guess values for $\Gamma$ and $k$. Since the standing wave is added into the received signal, it is then subtracted (in dB) from the from the path loss formula in equation \ref{sffl} to obtain the final model. The least squares linear fit is repeated with $\Gamma$ and $k$ as optimizing variables to provide more accurate fit with smaller root mean square value compared to fitting solely the {\em single-frequency floating intercept model}.

\subsection{Parameter Estimation for Phase} \label{phase}
 The phase shift according to frequency change is shown in equation \eqref{phaseEq}
 
\begin{equation}
 \Delta\Phi (rad) = \frac{2\pi d \Delta f}{c}
\label{phaseEq}
\end{equation}
where $\Delta\Phi$ is the phase shift (rad), d is the distance, c is the speed of light, and $\Delta f$ is the change in frequency.

\begin{figure}[th]
  \centerline
      {\includegraphics[width=3.5in]{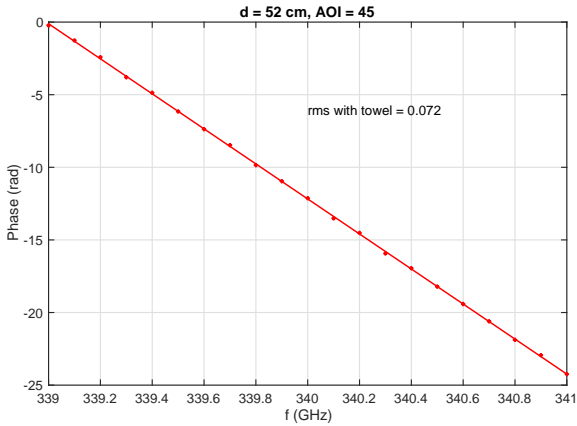}}
      \caption{Linear phase fit with d = 52 cm, $f_c$ = 340 GHz at $45^o$ AOI}
      \label{singleFit}
\end{figure}

First, we apply the linear fit to a small range of frequency around the interested frequency to obtain the fitted distance from the slope of the measured data. Figure \ref{singleFit} shows the fitting for 340 GHz center frequency at d = 52 cm. The phase response is a straight linear line shows that the system is correctly calibrated. However, the distance calculated from the slope is 57.55 cm. This is different from the measured distance of 52 cm even if we add 1.68cm x 2 (two times the antennas length). The other fitted distances at 340 GHz are listed in Table \ref{table2}.

\begin{table}[h]
\centering
\caption{Distance (in cm) from phase linear fitting}
\begin{tabular}{|c|cc|cc|} \hline
\setlength{\tabcolsep}{40pt}
\multirow{2}{4em}{d$_{meas}$} & \multicolumn{2}{|c|}{340 GHz} & \multicolumn{2}{|c|}{480 GHz}\\
& d$_{fit}$& $\Delta d$ & d$_{fit}$ & $\Delta d$ \\ \hline\hline
30.40   & 35.90  & 5.50     & 33.92 & 3.52 \\
34.00   & 39.50  & 5.50     & 37.49 & 3.49 \\
37.60   & 43.28  & 5.68     & 41.24 & 3.64 \\
41.20   & 46.67  & 5.47     & 44.52 & 3.32 \\
44.80   & 50.34  & 5.54     & 48.07 & 3.27 \\
48.40   & 53.62  & 5.22     & 51.64 & 3.24 \\
52.00   & 57.55  & 5.55     & 55.50 & 3.50 \\
55.60   & 61.02  & 5.42     & 58.81 & 3.21 \\
59.20   & 64.62  & 5.42     & 62.45 & 3.25 \\
62.80   & 68.26  & 5.46     & 66.27 & 3.47 \\
66.40   & 72.02  & 5.62     & 69.90 & 3.50 \\
70.00   & 75.26  & 5.26     & 73.37 & 3.37\\ \hline
\end{tabular}
\label{table2}
\end{table}

\begin{figure}[th]
  \centerline
      {\includegraphics[width=3.5in]{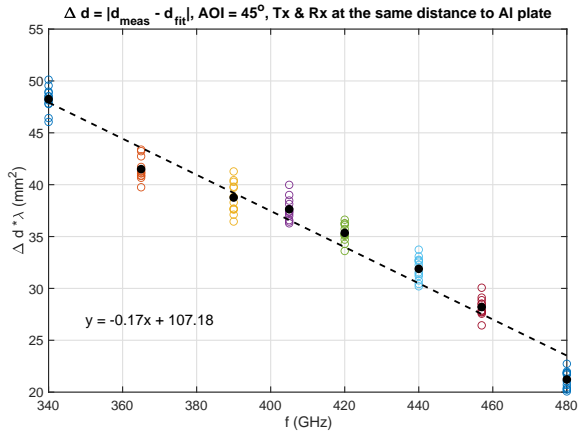}}
      \caption{Linear fitting of $\Delta d*\lambda$ at different frequency at $45^o$ AOI. Colored circles are from linear phase fitting at different distances. Black circles are the average values for a given frequency.}
      \label{dError}
\end{figure}

To examine this discrepancy between measured distance and fitted distance, we apply the same linear fit for the rest of the distances and center frequencies (410, 460 GHz). We observe that {\em the difference between measured and fitted distance remains constant for a given frequency}. Furthermore, the product of the difference in distance $\Delta d$ and the wavelength yields a linear function vs frequency as shown in Figure \ref{dError}. The linear equation from Figure \ref{dError} can be used to predict the error in distance $\Delta d$ for a given frequency.
\begin{equation}
    \Delta d * \lambda = \alpha*f + \beta
    \label{predict_delta_d}
\end{equation}
where $\Delta d$ is the difference between measured and fitted data, $\lambda$ is wavelength and $f$ is frequency.

Equation \ref{predict_delta_d} can thus be used for phase prediction for a given frequency and distance. The algorithm to do this is:
\medskip

\begin{enumerate}
    \item Choose a distance $d$ and a center frequency $f$.
    \item Use Equation \ref{predict_delta_d} to find the additional distance $\Delta d$ needed to add to distance $d$.
    \item With the corrected distance, use Equation \ref{phaseEq} to calculate the phase of the transmitted signal.
\end{enumerate}
\medskip

\section{Measurements and Validation} \label{validate} 

We conducted measurements for three frequencies: 340, 410, and 460 GHz. For each, we took multiple measurements at 12 distances (total path length from the Tx to metal/wood plate to Rx). Figures \ref{Magfitting340GHz},\ref{Magfitting410GHz},\ref{Magfitting460GHz} plot the attenuation of the received signals against distance for each case. Observe that the dots (which are the measured data) show increasing attenuation with distance, as expected, but also a reduction in attenuation for some distances. This is caused by standing waves that get established between the Tx and Rx, as we described in \cite{tranEffectStandingWave2021}. It is noteworthy that the standing wave exists even when the signal is reflected. The next next two sections we describe our fit for amplitude and phase.

\subsection{Magnitude fitting}

\begin{figure}[th]
  \centerline
      {\includegraphics[width=3.5in]{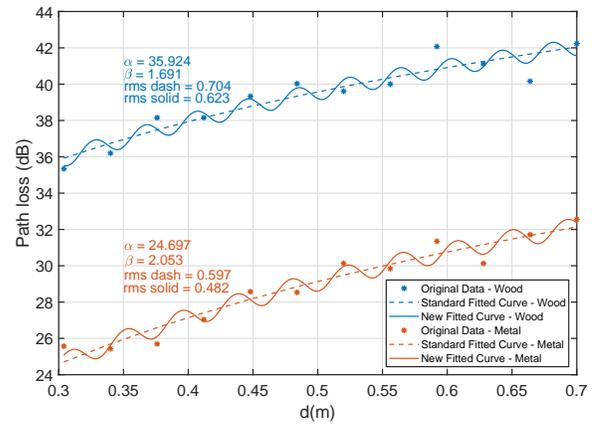}}
      \caption{Magnitude fit with reflected signal from wooden board (blue) and metal plate (red) for 340 GHz at $45^o$ AOI.}
      \label{Magfitting340GHz}
\end{figure}
 
 \begin{figure}[th]
  \centerline
      {\includegraphics[width=3.5in]{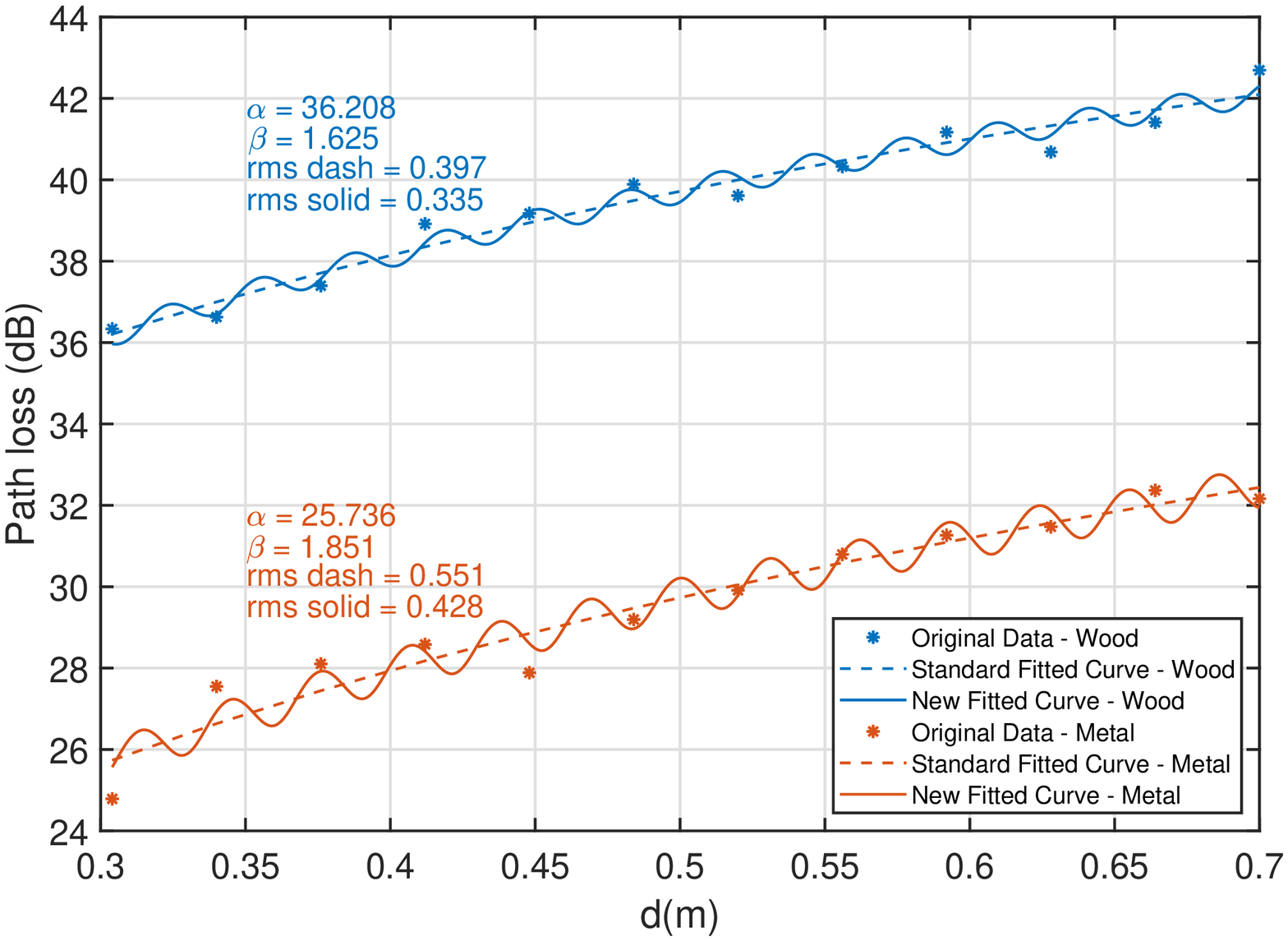}}
      \caption{Magnitude fitting with reflected signal from wooden board (blue) and metal plate (red) for 410 GHz at $45^o$ AOI.}
      \label{Magfitting410GHz}
\end{figure}

\begin{figure}[th]
  \centerline
      {\includegraphics[width=3.5in]{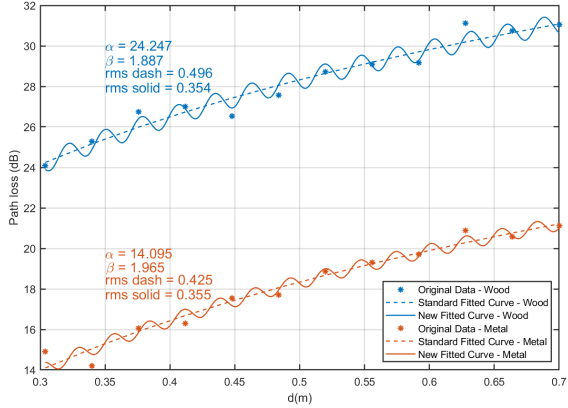}}
      \caption{Magnitude fitting with reflected signal from wooden board (blue) and metal plate (red) for 460 GHz at $45^o$ AOI.}
      \label{Magfitting460GHz}
\end{figure}

The path loss fitting of frequencies 340 GHz, 410 GHz, 460 GHz for both metal plate and wooden board as reflection surface are illustrated in Figures \ref{Magfitting340GHz},\ref{Magfitting410GHz},\ref{Magfitting460GHz}. As expected, the fitting with the standing wave effect give an improved agreement with the measured data. {\em Overall, the path loss in the metal plate case is around 10dB lower than the measurements with the wooden board.} The root mean square values are reasonable compared to the line of sight path loss cases in \cite{tranEffectStandingWave2021}.

The reflection coefficient $\Gamma$ and wavenumber $k$ from the fitting is listed in Table \ref{table3}. The standing wave phenomena still holds in this NLoS case as metal acts as an almost perfect reflector as this frequency range \cite{naftalyTerahertzReflectivitiesMetalcoated2011}.

\begin{table}[]
\centering
\caption{Magnitude fitting parameters}
\begin{tabular}{|l|l|l|l|}\hline
         & 340 GHz                 & 410 GHz                 & 460 GHz                 \\ \hline
         \multicolumn{4}{|c|}{Metal Plate} \\\hline
$|\Gamma|$ & $0.06$   & $0.06$  & $0.04$  \\
$\angle\Gamma$ & $1.47\pi$   & $-0.23\pi$  & $-0.18\pi$  \\
k        & $23.49\pi$ & $33.85\pi$ & $36.75\pi$ \\ \hline
         \multicolumn{4}{|c|}{Wood Plate} \\\hline
$|\Gamma|$ & $0.05$  & $0.03$  & $0.06$  \\
$\angle\Gamma$ & $1.37\pi$ & $-0.35\pi$ & $1.19\pi$ \\
k        & $25.42\pi$ & $31.49\pi$ & $35.47\pi$  \\ \hline
\end{tabular}
\label{table3}
\end{table}

We observe that the reflection coefficient $|\Gamma|$ for the standing wave is slightly different when comparing the LoS case from \cite{tranEffectStandingWave2021} as opposed to the NLoS case studied in this paper, Table \ref{los-nlos}. This is because reflections from the metal plate affects the overall measured value for $\Gamma$. Thus, we have multiple standing waves established between Tx-plate, Tx-plate-Rx, plate-Rx, Tx-plate-Rx-plate-Rx, and some others. The value reported in Table \ref{los-nlos} is the cumulative effect of all these standing waves. Observe that the presence of these also affect the measured phase. However, since the amplitude is relatively small, these cause only a small phase error.
\begin{table}[h]
\centering
\caption{Comparison of reflection coefficient (NLoS uses Metal Plate).}
\label{los-nlos}
\begin{tabular}{|c|c|c|}\hline
$f$ & LoS & NLoS \\
& $|\Gamma|$ & $|\Gamma|$\\ \hline
340 & 0.05 & 0.06\\
410 & 0.09 & 0.06\\
460 & 0.06 & 0.04\\ \hline
\end{tabular}
\end{table}

\subsection{Phase fitting}

\begin{figure}[th]
  \centerline
      {\includegraphics[width=3.5in]{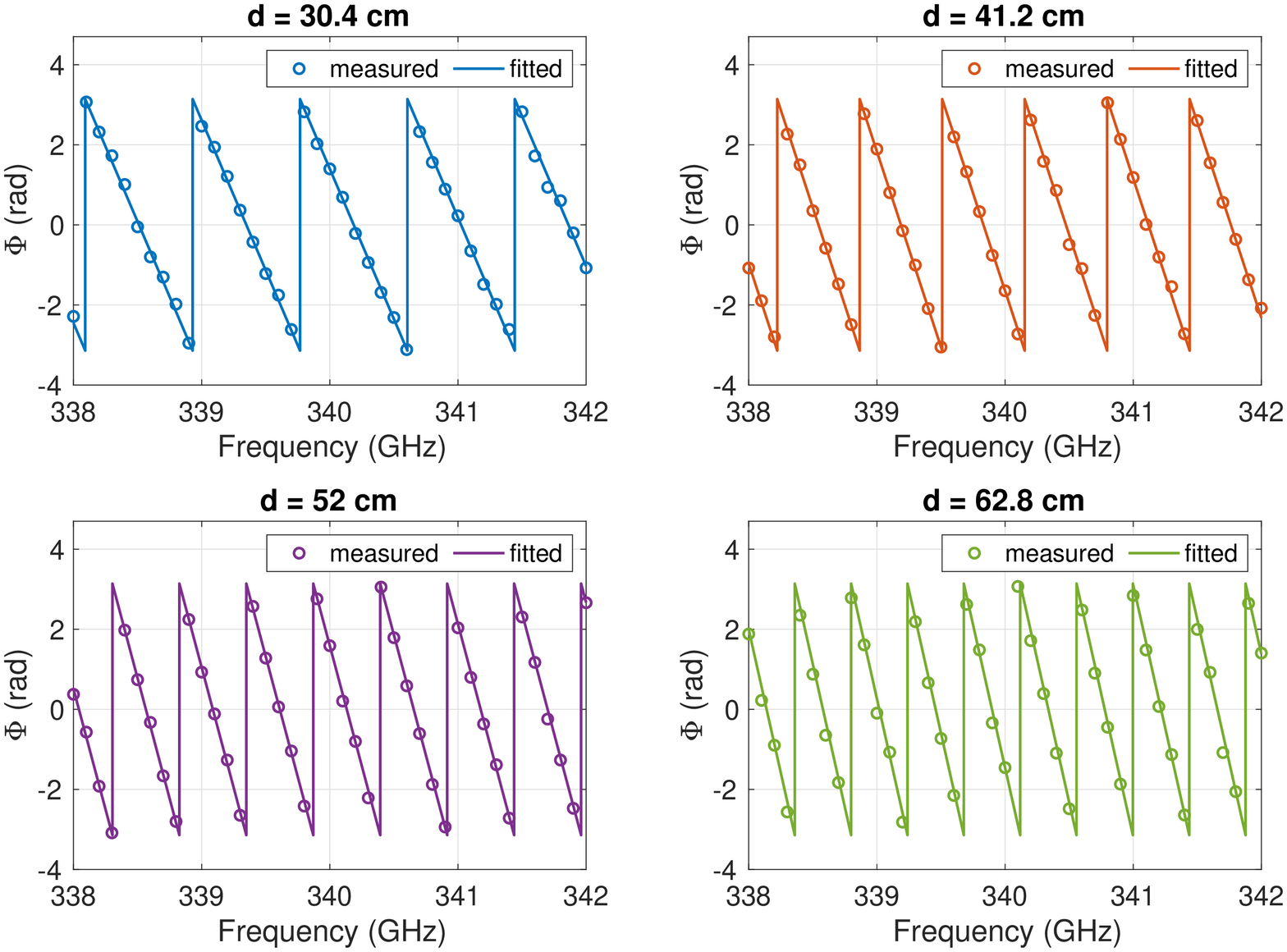}}
      \caption{Phase fitting at $f_c = 340$ GHz at $45^o$ AOI for distances of 30.4, 41.2, 52, and 62.8 cm.}
      \label{Phasefitting340GHz}
\end{figure}

We use the algorithm outlined in section \ref{phase} to fit the phase for each of the three center frequencies. For each frequency, we first compute $\Delta d$ (distance between measured and fitted distance) and correct the measured distance by adding $\Delta d$. This fitting yields:

\begin{equation}
    \Delta d(m) * \lambda(m) = -0.17*f (GHz) + 107.18
    \label{predict_delta_d_validate}
\end{equation}
This corrected distance is then used to compute the phase from equation \ref{phaseEq}. Phase around the center frequency 340 GHz at distances of 30.4, 41.2, 52, 62.8 cm is plotted in Figure \ref{Phasefitting340GHz}. We obtain an excellent match between the measured and predicted phases. Other measurements around 410 GHz, 460 GHz also gives the same quality match. The measurements using wooden board for these frequencies show a similar linear relationship to \eqref{predict_delta_d_validate} with slope of -0.18 and offset of 106.68. We obtain very good agreement between the fitted and measured phase data. 

\section{Conclusion} \label{conclude}
 In this paper, a complete channel model is obtained and verified for the case of a single reflection for terahertz when the reflection is off a metal plate and a wooden board. The amplitude fit, with the inclusion of standing wave, is much better compared to the standard single frequency floating intercept model. We have also provided an algorithm for computing the parameters of the fit for the magnitude. 
 
 In addition, we have developed a model to predict the phase of the received signal after a single reflection. We observed that for each frequency, there is a constant discrepancy between the actual Tx-wall-Rx distance and the distance predicted by using the arriving phase. We observe that this discrepancy in predicted vs actual distance has a linear relationship to frequency. Using this observation, we have developed an algorithm to correct the predicted phase at the receiver and we show that our model provides a phase that is very close to actual measured phase.
 
 In current and future work we are extending the model to reflections from other materials typically used in indoor spaces. In addition, we are developing models for multiple reflections for frequencies upto 520 GHz.
 
\bibliographystyle{unsrt}
\bibliography{UsableReferences, Usable, OurPapers, thz, reference, 5g, others}

\end{document}